\documentclass[a4paper,14pt]{article}
\usepackage[english]{babel}
\pdfoutput=1 

\usepackage{jheppub} 

\usepackage[T1]{fontenc} 

\usepackage[T1,T2A]{fontenc}
\usepackage[utf8x]{inputenc}

\usepackage{comment}

\usepackage{cases}

\newcommand{\bi}{\begin{itemize}}
	\newcommand{\ei}{\end{itemize}}
\newcommand{\bea}{\begin{eqnarray}}
	\newcommand{\eea}{\end{eqnarray}}
\newcommand{\bt}{\begin{tabular}}
	\newcommand{\et}{\end{tabular}}
\newcommand{\bc}{\begin{center}}
	\newcommand{\ec}{\end{center}}

\newcommand{\be}{\begin{equation}}
	\newcommand{\ee}{\end{equation}}
\newcommand{\ba}{\begin{array}}
	\newcommand{\ea}{\end{array}}

\newcommand{\lb}[1]{\label{#1}}

\def\bbox{{\,\lower0.9pt\vbox{\hrule \hbox{\vrule height 0.2 cm
				\hskip 0.2 cm \vrule height 0.2 cm}\hrule}\,}}
\newcommand{\dsl}{\pa \kern-0.5em /}

\makeatletter \@addtoreset{equation}{section} \makeatother

\def\slashchar#1{\setbox0=\hbox{$#1$}           
	\dimen0=\wd0                                 
	\setbox1=\hbox{/} \dimen1=\wd1               
	\ifdim\dimen0>\dimen1                        
	\rlap{\hbox to \dimen0{\hfil/\hfil}}      
	#1                                        
	\else                                        
	\rlap{\hbox to \dimen1{\hfil$#1$\hfil}}   
	/                                         
	\fi}

\pdfoutput=1

\title{\boldmath $\mathcal{N}=2$ higher-spin supercurrents}

\author[a,b]{Nikita~Zaigraev}

\affiliation[a]{Bogoliubov Laboratory of Theoretical Physics, JINR,\\141980 Dubna, Moscow region, Russia}
\affiliation[b]{Moscow Institute of Physics and Technology,\\ 141700 Dolgoprudny, Moscow region, Russia}

\emailAdd{nikita.zaigraev@phystech.edu}

\abstract{We have constructed conserved  $\mathcal{N}=2$ higher-spin supercurrents for an arbitrary integer spin through the unconstrained superfield formulation of $4D,\, \mathcal{N}=2$ massless higher-spin theories in the harmonic superspace. The obtained supercurrents are gauge-invariant and determine consistent cubic interactions in the \textit{gauge superfield} $\times$ \textit{supercurrent} form for massless spin $\mathbf{s_1}$ and two massless spin $\mathbf{s_2}$ $\mathcal{N}=2$ supermultiplets. Such interactions and $\mathcal{N}=2$ supercurrents exist only for $\mathbf{s_1}\geq 2 \mathbf{s_2}$. These supercurrents are the $\mathcal{N}=2$ supersymmetric extension of Berends-Burgers-van Dam higher-spin currents and generalize the linearized Bel-Robinson tensor. The relevant $\mathcal{N}=2$ supercurrents can be considered as the descendants of the $\mathcal{N}=2$ principle supercurrent, which has a simple universal structure. An important feature of our outcomes is that, ultimately, all $\mathcal{N}=2$ supercurrents are built from the $\mathcal{N}=2$ superfield strengths, which are constructed from the unconstrained analytical higher-spin prepotentials.}

\makeatletter
\gdef\@fpheader{}
\makeatother

\begin{document} 
\maketitle
\flushbottom

\section{Introduction}

Supercurrent \cite{Ferrara:1974pz, Ogievetsky:1976qc}
is the fundamental object in the supersymmetric theories. The supercurrent multiplet contains the energy-momentum tensor and conserved currents corresponding to supersymmetric invariance and acts as the source for supergravity.
$\mathcal{N}=2$ supercurrents for the hypermultiplet and $\mathcal{N}=2$ vector multiplet are given~\cite{Howe:1981qj, Kuzenko:1999pi, Butter:2010sc} by
\begin{equation}\label{eq: N=2 gr SC}
	\mathcal{J}_{hyp} = q^{+a} \mathcal{D}^{--} q^+_a,
	\qquad\qquad
	\mathcal{J}_{vec} = \mathcal{W} \bar{\mathcal{W}}
\end{equation}	
and satisfy the on-shell conservation equations\footnote{Here and in the following, we use the “weak equality” symbol $\approx$ to denote equations satisfied on the equations of motion.} in the harmonic superspace using notations of \cite{18}:
\vspace{-0,2 cm}
\begin{equation}
	\begin{cases}
		\mathcal{D}^{++} \mathcal{J}_{hyp} \approx 0,
		\\
		(\mathcal{D}^+)^2 \mathcal{J}_{hyp} = 0,
		\\
		(\bar{\mathcal{D}}^+)^2 \mathcal{J}_{hyp} = 0,
	\end{cases}	
	\qquad
		\begin{cases}
		\mathcal{D}^{++} \mathcal{J}_{vec} = 0,
	\\
	(\mathcal{D}^+)^2 \mathcal{J}_{vec} \approx 0,
	\\
	(\bar{\mathcal{D}}^+)^2 \mathcal{J}_{vec} \approx 0.
	\end{cases}	
\end{equation}	
Higher-spin gauge-invariant\footnote{For the discussion of gauge non-invariant higher-spin currents, see, e.g., \cite{Smirnov:2013kba}.} conserved currents exist for $s_1\geq 2s_2$ \cite{Berends:1985xx} and can be represented as\footnote{We use shortened notation for symmetrized product of derivatives $\partial^p := \partial_{\alpha(p)\dot{\alpha}(p)}^p = \partial_{(\alpha_1(\dot{\alpha}_1}\dots \partial_{\alpha_p)\dot{\alpha}_p)}$ and assume symmetrization over all doted and undoted indices.}:
\vspace{-0,2 cm}
\begin{equation}\label{eq: N=0 current}
	J_{\alpha(s_1)\dot{\alpha}(s_1)} 
	=
	c \, (i)^{s_1-2s_2} \sum_{p=0}^{s_1-2s_2} (-1)^p \frac{\binom{s_1}{2s_2+p}\binom{s_1}{p}}{\binom{s_1}{2s_2}}\;  \partial^p C_{\alpha(2s_2)}\,
	\partial^{(s_1-2s_2-p)} \bar{C}_{\dot{\alpha}(2s_2)},
\end{equation}	
where we used notation for the higher-spin generalization of the Weyl tensor:
\vspace{-0,2 cm}
\begin{equation}\label{eq: bos C}
	C_{(\alpha_1\dots\alpha_{2s})} = \partial_{(\alpha_1}^{\;\;\dot{\beta}_1} \dots \partial_{\alpha_s}^{\;\;\dot{\beta}_s} \Phi_{\alpha_{s+1}\dots \alpha_{2s})(\dot{\beta}_1 \dots \dot{\beta}_s)},
	\qquad
		\partial_{\dot{\alpha}}^{\;\;\alpha_1} 	C_{(\alpha_1\dots\alpha_{2s})}  \approx 0.
\end{equation}	
In the $s=1$ case, this tensor corresponds to the self-dual part of the electromagnetic tensor, and in the $s=2$ case, to the linearized Weyl tensor's self-dual part.
Higher-spin currents \eqref{eq: N=0 current} satisfy the on-shell conservation equation:
\begin{equation}\label{eq: bos cons eq}
	\partial^{\alpha\dot{\alpha}} J_{\alpha(s_1)\dot{\alpha}(s_1)}  \approx 0.
\end{equation}	
These conserved currents are higher-spin generalizations of the linearized gravitational Bel-Robinson tensor \cite{Bel}, which is the well-known spin $4$ conserved current in general relativity.

Using conserved higher-spin currents \eqref{eq: N=0 current},
one can construct the consistent cubic
$(s_1, s_2, s_2)$ vertex \cite{Berends:1985xx,Deser:1990bk, Manvelyan:2009vy} for $s_1\geq 2s_2$
\begin{equation}\label{eq: type 2}
	\mathcal{L}^{(s_1, s_2, s_2)}_{int} \sim \Phi^{\alpha(s_1)\dot{\alpha}(s_1)} J_{\alpha(s_1)\dot{\alpha}(s_1)}.
\end{equation}	


In this article, we consider the problem of the construction of $\mathcal{N}=2$ higher-spin supercurrents for an arbitrary integer higher spin\footnote{The $\mathcal{N}=1$ supersymmetric extension of gauge-invariant higher-spin currents \eqref{eq: N=0 current} and the corresponding cubic vertices were constructed in \cite{Buchbinder:2018wzq}. Another remarkable class of conserved $\mathcal{N}=1$ supercurrents and corresponding consistent non-minimal cubic interactions were considered in the paper \cite{Gates:2019cnl}.} and the corresponding Abelian cubic interactions. The basis for our construction is the harmonic superspace formulation~\cite{Galperin:1984av, 18} of the $\mathcal{N}=2$ theory of higher spins \cite{Buchbinder:2021ite} in terms of unconstrained analytical prepotentials. Based on these results, it was shown that off-shell higher-spin hypermultiplet cubic vertices and corresponding hypermultiplet $\mathcal{N}=2$ supercurrents have an elegant structure \cite{Buchbinder:2022kzl, Buchbinder:2022vra, Buchbinder:2024pjm}; see also reviews \cite{Zaigraev:2023ogo, Zaigraev:2024xve}. However, more general higher-spin Abelian vertices and $\mathcal{N}=2$ supercurrents were never considered. Some general properties of $\mathcal{N}=2$ supercurrents were analyzed by Kuzenko and Raptakis in \cite{Kuzenko:2023vgf}, settled on their earlier paper on the $\mathcal{N}${--}extended superconformal multiplets. \cite{Kuzenko:2021pqm}. In particular, the authors established that on-shell the analytical hypermultiplet higher-spin $\mathcal{N}=2$ supercurrents can be considered as the descendants of the $\mathcal{N}=2$ superconformal primary supercurrent and also constructed $\mathcal{N}=2$ primary higher-spin supercurrents for the $\mathcal{N}=2$ vector multiplet\footnote{At the final stage of preparation for this work, the preprint by Kuzenko and Raptakis \cite{Kuzenko:2024vms} appeared, where authors constructed $\mathcal{N}=2$ primary higher-spin supercurrents for  the off-shell hypermultiplet. }.

Our goal is to continue the analysis of $\mathcal{N}=2$ supercurrents based on the harmonic superspace approach and explicitly construct gauge-invariant $\mathcal{N}=2$ higher-spin supercurrents in terms of analytical prepotentials for an arbitrary integer spin.
For the construction of gauge-invariant $\mathcal{N}=2$ higher-spin supercurrents, the main objects are higher-spin superfield strengths; those provide the manifest gauge invariance. In the recent paper~\cite{Ivanov:2024gjo}, the $\mathcal{N}=2$ super-Weyl tensor in terms of the $\mathcal{N}=2$ Einstein supergravity analytic prepotentials was obtained.  In this work, we intend to generalize the $\mathcal{N}=2$ super-Weyl tensor to the higher spins and then use it to construct conserved $\mathcal{N}=2$ higher-spin supercurrents for an arbitrary integer spin.

\vspace{-0,2 cm}

\section{$\mathcal{N}=2$ massless higher-spin theories in harmonic superspace}\label{sec: 2}

The massless off-shell $\mathcal{N}=2$  \textit{spin $\mathbf{s}\geq2$ supermultiplet} is described by the set of unconstrained analytic prepotentials \cite{Buchbinder:2021ite}:
\vspace{-0,2 cm}
\begin{equation}\label{eq: HS analytic prepotentials}
	h^{++\alpha(s-1)\dot{\alpha}(s-1)},
	\quad
	h^{++\alpha(s-1)\dot{\alpha}(s-2)+},
	\quad
	h^{++\alpha(s-2)\dot{\alpha}(s-1)+},
	\quad
	h^{++\alpha(s-2)\dot{\alpha}(s-2)}.
	\vspace{-0,3 cm}
\end{equation}	
The gauge transformations of the analytic prepotentials are the following:
 \begin{equation}\label{eq: GF}
	\begin{split}
		\delta_\lambda h^{++\alpha(s-1)\dot\alpha(s-1)} &= \mathcal{D}^{++} \lambda^{\alpha(s-1)\dot\alpha(s-1)} 
		\\&
		\qquad +
		4i \big[\lambda^{+\alpha(s-1)(\dot\alpha(s-2)}\bar\theta^{+\dot\alpha_{s-1})} +\theta^{+(\alpha_{s-1}} \bar\lambda^{+\alpha(s-2))\dot\alpha(s-1)} \big], 
		\\
		\delta_\lambda h^{++\alpha(s-2)\dot\alpha(s-2)} & =  \mathcal{D}^{++} \lambda^{\alpha(s-2)\dot\alpha(s-2)}
		\\&
		\qquad -
		2i\,\big[\lambda^{+(\alpha(s-2)\alpha_{s-1})\dot\alpha(s-2)} \theta^+_{\alpha_{s-1}} 
		+
		\bar\lambda^{+(\dot\alpha(s-2)\dot\alpha_{s-1})\alpha(s-2)} \bar\theta^+_{\dot\alpha_{s-1}} \big], 
		\\
		\delta_\lambda  h^{++\alpha(s-1)\dot\alpha(s-2)+} &=  \mathcal{D}^{++} \lambda^{+\alpha(s-1)\dot\alpha(s-2)}\,,
		\\
		\delta_\lambda h^{++\dot\alpha(s-1)\alpha(s-2)+} &=
		\mathcal{D}^{++} \bar\lambda^{+\dot\alpha(s-1)\alpha(s-2)}. 
	\end{split}
\end{equation}
The realization of rigid $\mathcal{N}=2$ supersymmetry transformations on analytic  higher-spin prepotentials is given by equations 
 \begin{equation}
	\begin{split}
		\delta_\epsilon h^{++\alpha(s-1)\dot\alpha(s-1)} &= -4i\big[h^{++\alpha(s-1)(\dot\alpha(s-2)+}\bar\epsilon^{-\dot\alpha_{s-1})}-
		h^{++\dot\alpha(s-1)(\alpha(s-2)+}\,\epsilon^{-\alpha_{s-1})}
		\big]\,, 
		\\
		\delta_\epsilon h^{++\alpha(s-2)\dot\alpha(s-2)} &= 2i\big[h^{++(\alpha(s-2)\alpha_{s-1})
			\dot\alpha(s-2)+}\epsilon^{-}_{\alpha_{s-1}} +
		h^{++\alpha(s-2)(\dot\alpha(s-2)\dot\alpha_{s-1})+}\,\bar\epsilon^{-}_{\dot{\alpha}_{s-1}}
		\big]\,,
		\\
		\delta_\epsilon h^{++\alpha(s-1)\dot{\alpha}(s-2)+} &= 0,
		\\
		\delta_\epsilon h^{++\dot{\alpha}(s-1)\alpha(s-2) +} &= 0,
	\end{split}
\end{equation}
where we have used the notations $\epsilon^{\pm \alpha} := \epsilon^{\alpha i} u^\pm_i$ and $\bar{\epsilon}^{\pm \dot{\alpha}} := \bar{\epsilon}^{ \dot{\alpha} i} u^\pm_i $.

$\mathcal{N}=2$ supersymmetry invariant superfields  are defined by relations:
\begin{equation}\label{eq: G++ def}
	\begin{split}
		G^{++\alpha(s-1)\dot\alpha(s-1)} & =
		h^{++\alpha(s-1)\dot\alpha(s-1)} + 4i
		\big[h^{++\alpha(s-1)(\dot\alpha(s-2)+}\bar\theta^{-\dot\alpha_{s-1})}
		\\
		&\qquad\qquad\qquad\qquad\qquad\quad -
		h^{++\dot\alpha(s-1)(\alpha(s-2)+}\,\theta^{-\alpha_{s-1})}
		\big], \\ 
        G^{++\alpha(s-2)\dot\alpha(s-2)} &=
		h^{++\alpha(s-2)\dot\alpha(s-2)} - 2i
		\big[h^{++(\alpha(s-2)\alpha_{s-1})
			\dot\alpha(s-2)+}\theta^{-}_{\alpha_{s-1}} 
				\\&\qquad\qquad\qquad\qquad\qquad\quad +
		h^{++\alpha(s-2)(\dot\alpha(s-2)\dot\alpha_{s-1})+}\,\bar\theta^{-}_{\dot{\alpha}_{s-1}}
		\big],
		\\
		G^{++\alpha(s-1)\dot\alpha(s-2)+} &= - h^{++\alpha(s-1)\dot\alpha(s-2)+}, 
		\\
		G^{++\alpha(s-2)\dot\alpha(s-1)+} &= - h^{++\alpha(s-2)\dot\alpha(s-1)+}.
	\end{split}
\end{equation}
Bosonic $G^{++\dots}$ potentials are not analytical but supersymmetry-invariant.
Through the use of the zero-curvature conditions\footnote{We use notations for indices $M = \{\alpha\dot{\alpha}, \alpha, \dot{\alpha}, 5 \}$ and $\hat{\alpha} = \{\alpha, \dot{\alpha} \}$. We also omit the index 5, so $h^{++\alpha(s-2)\dot{\alpha}(s-2)5} = h^{++\alpha(s-2)\dot{\alpha}(s-2)}$.}
  \begin{equation}\label{zero-curv}
  	\begin{split}
  	 \mathcal{D}^{++}G^{--\alpha(s-2)\dot\alpha(s-2)M} &= \mathcal{D}^{--}G^{++\alpha(s-2)\dot\alpha(s-2)M}\,, 
  	 \\
  	 \mathcal{D}^{++}G^{--\alpha(s-2)\dot{\alpha}(s-2)\hat{\beta}-}
  	 &+
  	 G^{--\alpha(s-2)\dot{\alpha}(s-2)\hat{\beta}+} = 0,
  	\end{split} 
  \end{equation}
one can construct supersymmetry-invariant negatively charged potentials $G^{--\dots}$, which play an important role in the harmonic superfield formulation of $\mathcal{N}=2$ theories.

  Gauge invariant $\mathcal{N}=2$ supersymmetric action has the universal form for all spin $\mathbf{s}$  supermultiplets \cite{Buchbinder:2021ite}:
  \begin{equation}\lb{ActionsGen}
  	\begin{split}
  		& S_{(s)} = (-1)^{s+1} \int d^4x
  		d^8\theta du \,\Big\{G^{++
  			\alpha(s-1)\dot\alpha(s-1)}G^{--}_{\alpha(s-1)\dot\alpha(s-1)} \\
  		&\;\;\;\;\;  \qquad\qquad \qquad\qquad \qquad\qquad+\,
  		4G^{++\alpha(s-2)\dot\alpha(s-2)}G^{--}_{\alpha(s-2)\dot\alpha(s-2)}
  		\Big\}\,. 
  	\end{split}
  \end{equation}
  By varying this action with respect to the analytic prepotentials, one can derive  superfield equations of motion \cite{Buchbinder:2022vra}
  \begin{equation}\label{eq: super EOM}
  	(\bar{\mathcal{D}}^+)^2 \mathcal{D}^{+\alpha_{s-1}} G^{--}_{\alpha(s-1) \dot{\alpha}(s-1)}
  	-4
  	(\mathcal{D}^+)^2 \bar{\mathcal{D}}^+_{(\dot{\alpha}} G^{--}_{\alpha(s-2) \dot{\alpha}(s-2))} \;
  	\approx 0.
  \end{equation}
  In components, these equations lead to the free equations for physical gauge fields. Auxiliary fields are equal to zero on-shell (for $\mathbf{s}\geq3$).  
  
  The gauge-invariant $\mathcal{N}=2$ spin $\mathbf{s}$ superfield strength is defined as
   \begin{equation}\label{eq: SF strenghts}
  	\begin{split}
  		\mathcal{W}_{\alpha(2s-2)}
  		:=
  		&\left(\bar{\mathcal{D}}^+\right)^2 \partial_{(\alpha_1}^{\;\;\dot{\beta}_1} \dots  	\partial_{\alpha_{s-2}}^{\;\;\dot{\beta}_{s-2}} \Bigr\{
  		\mathcal{D}^+_{\alpha_{s-1}} G^{---}_{\alpha_s\dots\alpha_{2s-2})(\dot{\beta}_1\dots\dot{\beta}_{s-2})}
  		\\&\;\;\;	+
  		\mathcal{D}^-_{\alpha_{s-1}}
  		G^{--+}_{\alpha_s\dots\alpha_{2s-2})(\dot{\beta}_1\dots\dot{\beta}_{s-2})}
  		-
  		\partial_{\alpha_{s-1}}^{\;\;\dot{\beta}_{s-1}} G^{--}_{\alpha_s\dots\alpha_{2s-2})(\dot{\beta}_1\dots\dot{\beta}_{s-1})}\Bigr\}.
  	\end{split}
  \end{equation}	
This is the generalization of the spin $\mathbf{2}$ super-Weyl tensor, introduced in the harmonic superspace in recent work \cite{Ivanov:2024gjo}. Analogously, one can define conjugated superfield strength $	\bar{\mathcal{W}}_{\dot{\alpha}(2s-2)} := \widetilde{\mathcal{W}}_{\alpha(2s-2)}$.
  
  The superstrength $\mathcal{W}_{\alpha(2s-2)}$ is the chiral $\mathcal{N}=2$ superfield, i.e., $\bar{\mathcal{D}}^\pm_{\dot{\beta}}\mathcal{W}_{\alpha(2s-2)} = 0$ and the tilde conjugated strength $\bar{\mathcal{W}}_{\dot{\alpha}(2s-2)}$ is the anti-chiral superfield, i.e., $\mathcal{D}^\pm_{\beta} \bar{\mathcal{W}}_{\dot{\alpha}(2s-2)} = 0$. Moreover, $\mathcal{W}_{\alpha(2s-2)}$ and $\bar{\mathcal{W}}_{\dot{\alpha}(2s-2)}$ are covariantly harmonic-independent and satisfy the following equations
  \begin{equation}
  	\mathcal{D}^{\pm\pm} \mathcal{W}_{\alpha(2s-2)} = 0,
  	\qquad
  		\mathcal{D}^{\pm\pm} \bar{\mathcal{W}}_{\dot{\alpha}(2s-2)} = 0,
  \end{equation}	
   while on-shell superfield strengths satisfy
  \begin{equation}\label{eq: SEOM}
  	\mathcal{D}^{\pm\alpha}	\mathcal{W}_{\alpha(2s-2)} \approx 0,
  	\qquad
  	\bar{\mathcal{D}}^{\pm\dot{\alpha}}	\bar{\mathcal{W}}_{\dot{\alpha}(2s-2)}  \approx 0,
  	\qquad
  	\partial_{\dot{\beta}}^{\;\alpha } \mathcal{W}_{\alpha(2s-2)} \approx 0,
  	\qquad
  	\partial_{\beta}^{\;\dot{\alpha}} \bar{\mathcal{W}}_{\dot{\alpha}(2s-2)} \approx 0.
  \end{equation}	
  These equations follow from the component composition of the superfield strengths. Actually, since the superfield strengths are gauge-invariant, they contain higher-spin bosonic and fermionic  Weyl tensors in their component decomposition, which leads to the equations above. In particular, in components superfield strengths contain the Weyl-like higher spin tensor \eqref{eq: bos C}
  \begin{equation}
  	\mathcal{W}_{\alpha(2s-2)} \sim \theta^{+\beta} \theta^{-\gamma}  C_{(\alpha(2s-2)\beta\gamma)} + \dots
  \end{equation}	
  Thus, one can consider this $\mathcal{N}=2$ superfield strength as the natural higher spin generalization of the linearized $\mathcal{N}=2$ super-Weyl tensor.
  
  \medskip
  
  The $\mathcal{N}=2$ \textit{vector multiplet} can be considered as the limiting case of $\mathcal{N}=2$ higher spins. The off-shell $\mathcal{N}=2$ vector multiplet is described by the unconstrained analytic superfield $V^{++}$ with the gauge freedom 
  \begin{equation}\label{eq: spin 1 action1}
  	\delta_\lambda V^{++} = \mathcal{D}^{++} \lambda.
  \end{equation}	
  Gauge-invariant superfield action is given by:
  \begin{equation}\label{eq: spin 1 action}
  	S_{(s=1)} = \int d^4x d^8\theta du \; V^{++} V^{--};
  \end{equation}
  here we define $V^{--}$  as the solution of zero curvature equation
  	$ \mathcal{D}^{++} V^{--} = \mathcal{D}^{--} V^{++} $.
  The dynamical equations of motion can be derived from action \eqref{eq: spin 1 action}  
  \begin{equation}
  	(\mathcal{D}^+)^4 V^{--} \approx 0.
  \end{equation}	
  Gauge-invariant superfield strengths for the $\mathcal{N}=2$ vector multiplet are defined as follows:
  \begin{equation}\label{eq: W spin 1}
  	\mathcal{W} := (\bar{\mathcal{D}}^+)^2 V^{--},
  	\qquad
  	\bar{\mathcal{W}} := (\mathcal{D}^+)^2 V^{--}, 
  \end{equation}
  and, as their higher-spin counterparts, satisfy harmonic independence and chirality/antichirality conditions correspondingly.
   On-shell $\mathcal{W}$ and $\bar{\mathcal{W}}$ satisfy useful equations
  \begin{equation}\label{eq: spin 1 on-shell}
  		\left(\mathcal{D}^+\right)^2\mathcal{W}  \approx 0,
  		\qquad
  		\left(\bar{\mathcal{D}}^+\right)^2 	\bar{\mathcal{W}} \approx 0,
  		\qquad
  		\partial_{\dot{\alpha}}^{\;\alpha} \mathcal{D}^{\pm}_\alpha  \mathcal{W}  \approx 0,
  		\qquad
  		\partial_{\alpha}^{\;\dot{\alpha}} \bar{\mathcal{D}}^\pm_{\dot{\alpha}} \bar{\mathcal{W}} \approx 0
  		.
  \end{equation}

  \section{$\mathcal{N}=2$ higher spin Mezincescu prepotentials}\label{sec: 3}

 An alternative representation of $\mathcal{N}=2$ higher-spin prepotentials is given by the  Mezincescu-type higher-spin non-analytic prepotentials, introduced in \cite{Buchbinder:2022vra}. These prepotentials are natural harmonic superspace higher-spin generalizations of spin $\mathbf{1}$ Mezincescu  prepotential  \cite{Mezincescu} and spin $\mathbf{2}$ Gates-Siegel prepotential \cite{Gates:1981qq}.
 The most straightforward way to introduce Mezincescu-type prepotentials is to use 
 the supersymmetry-invariant analytical differential operator \cite{Buchbinder:2022kzl, Buchbinder:2022vra}
 \begin{equation}\label{eq: SUSY operator}
  	\hat{\mathcal{H}}^{++}_{(s)}   := h^{++\alpha(s-2)\dot{\alpha}(s-2)M} \partial_M \partial^{s-2}_{\alpha(s-2)\dot{\alpha}(s-2)},
 \end{equation}
  and represent it in the manifestly analytial form\footnote{Here we use notations $(\mathcal{D}^+)^4 := \frac{1}{16} (\mathcal{D}^+)^2 (\bar{\mathcal{D}}^+)^2$, $\mathcal{D}^+_{\hat{\alpha}} = \partial^+_{\hat{\alpha}}$, $\mathcal{D}^-_\alpha  = - \partial^-_\alpha + 4i \bar{\theta}^{-\dot{\alpha}} \partial_{\alpha\dot{\alpha}} $, $\bar{\mathcal{D}}^-_{\dot{\alpha}} = - \partial^-_{\dot{\alpha}} - 4i \theta^{-\alpha} \partial_{\alpha\dot{\alpha}}$.}
\begin{equation}\label{eq: Mez intr}
	\hat{\mathcal{H}}^{++}_{(s)}  
	=
	(\mathcal{D}^+)^4 \left[ \Psi^{-\alpha(s-1)\dot{\alpha}(s-2)} \mathcal{D}^{-}_{\alpha}
	+
	\bar{\Psi}^{-\alpha(s-2)\dot{\alpha}(s-1) } \bar{\mathcal{D}}^{-}_{\dot{\alpha}} \right] \partial^{s-2}_{\alpha(s-2)\dot{\alpha}(s-2)},
\end{equation}	
where the unconstrained superfields $\Psi^{-\alpha(s-1)\dot{\alpha}(s-2)}$ and $\bar{\Psi}^{-\alpha(s-2)\dot{\alpha}(s-1) }$ are spin $\mathbf{s}$ Mezincescu-type prepotentials.
Since the operator $\hat{\mathcal{H}}^{++}_{(s)} $ is supersymmetry-invariant $\delta_\epsilon \hat{\mathcal{H}}^{++}_{(s)} = 0$, under $\mathcal{N}=2$ supersymmetry we have
\begin{equation}
	\delta_\epsilon \Psi^{-\alpha(s-1)\dot{\alpha}(s-2)} = 0,
	\qquad
	\delta_\epsilon \bar{\Psi}^{-\alpha(s-2)\dot{\alpha}(s-1) } = 0.
\end{equation}
	
Equation \eqref{eq: Mez intr} defines an unambiguous relation of $\Psi^{-\dots}$ and $\bar{\Psi}^{-\dots}$ with the analytical prepotentials \eqref{eq: HS analytic prepotentials}. Relying on representation \eqref{eq: SUSY operator}, one can deduce the relation of the Mezincescu-type prepotential to the covariant $G^{++\alpha(s-2)\dot{\alpha}(s-2)M}$ superfields \eqref{eq: G++ def} as follows:
\begin{equation}\label{eq: G and Psi prepot}
	\begin{split}
		G^{++\alpha(s-1)\dot{\alpha}(s-1)} &=
		\frac{i}{2} \left( (\mathcal{D}^+)^2 \bar{\mathcal{D}}^{+(\dot{\alpha}} \Psi^{-\alpha(s-1)\dot{\alpha}(s-2))}
		+
		(\bar{\mathcal{D}}^+)^2 \mathcal{D}^{+(\alpha} \bar{\Psi}^{-\alpha(s-2))\dot{\alpha}(s-1)} \right),
	\\	
	G^{++\alpha(s-1)\dot{\alpha}(s-2)+} &= (\mathcal{D}^+)^4 \Psi^{-\alpha(s-1)\dot{\alpha}(s-2)},
	\\
		G^{++\alpha(s-2)\dot{\alpha}(s-1)+} &= (\mathcal{D}^+)^4 	\bar{\Psi}^{-\alpha(s-2)\dot{\alpha}(s-1) },
		\\
		G^{++\alpha(s-2)\dot{\alpha}(s-2)} &= \frac{i}{4} \left(  (\bar{\mathcal{D}}^+)^2 \mathcal{D}^+_\alpha \Psi^{-\alpha(s-1)\dot{\alpha}(s-2)} - (\mathcal{D}^+)^2 \bar{\mathcal{D}}^+_{\dot{\alpha}} \bar{\Psi}^{-\alpha(s-2)\dot{\alpha}(s-1)} \right).
	\end{split}
\end{equation}
This is the higher-spin generalization of alternative representations of $\mathcal{N}=2$ Einstein supergravity prepotentials, first introduced by Zupnik \cite{Zupnik:1998td}. 
 Higher-spin Mezincescu prepotentials are defined up to  the gauge freedom (parameters $K$) and  pre-gauge freedom (parameters $B$):
 \begin{subequations}\label{eq: Psi gauge and pre gauge}
\begin{equation}
	\begin{split}
		\delta_{\lambda,b} \Psi^{-}_{\alpha(s-1)\dot{\alpha}(s-2) }
		=
		\;
		&\mathcal{D}^{++} K^{(-3)}_{\alpha(s-1)\dot{\alpha}(s-2) }
		+
		\mathcal{D}^+_{(\alpha} B^{--}_{\alpha(s-2))\dot{\alpha}(s-2)}
		\\&+
		\mathcal{D}^{+\beta} B^{--}_{(\alpha(s-1)\beta)\dot{\alpha}(s-2)}
		+
		\bar{\mathcal{D}}^{+\dot{\beta}} B^{--}_{\alpha(s-1)(\dot{\alpha}(s-2)\dot{\beta})}
		\\
		&+
		\bar{\mathcal{D}}^+_{(\dot{\alpha}} B^{--}_{\alpha(s-1)\dot{\alpha}(s-3))},
	\end{split}
\end{equation}	
\begin{equation}
	\begin{split}
		\delta_{\lambda,b} \bar{\Psi}^{-}_{\alpha(s-2)\dot{\alpha}(s-1) }
		=\;
		&\mathcal{D}^{++} \bar{K}^{(-3)}_{\alpha(s-2)\dot{\alpha}(s-1) }
		-
		\bar{\mathcal{D}}^+_{(\dot{\alpha}} B^{--}_{\alpha(s-2)\dot{\alpha}(s-2))}
		\\&-
		\bar{\mathcal{D}}^{+\dot{\beta}} \bar{B}^{--}_{\alpha(s-2)(\dot{\alpha}(s-1)\dot{\beta})}
		+
		\mathcal{D}^{+\beta} B^{--}_{(\alpha(s-2)\beta)\dot{\alpha}(s-1)}
		\\&
		+\mathcal{D}^+_{(\alpha} \bar{B}^{--}_{\alpha(s-3))\dot{\alpha}(s-1)}.
		\end{split}
\end{equation}	
\end{subequations}
Parameters $K$ are related to the analytic $\lambda$-parameters through the relation: 
\begin{equation}
	\begin{split}
	\hat{\Lambda}_{(s)} &:=
	\lambda^{\alpha(s-2)\dot{\alpha}(s-2)M} \partial_M \partial^{s-2}_{\alpha(s-2)\dot{\alpha}(s-2)}
	\\
	&\, = (\mathcal{D}^+)^4 \left( K^{(-3)\alpha(s-1)\dot{\alpha}(s-2)} \mathcal{D}^-_\alpha
	+
	\bar{K}^{(-3)\alpha(s-2)\dot{\alpha}(s-1)} \bar{\mathcal{D}}^-_{\dot{\alpha}}
	 \right) \partial^{s-2}_{\alpha(s-2)\dot{\alpha}(s-2)}
	 \end{split}
\end{equation}	 
and define the gauge freedom \eqref{eq: GF} of the analytic prepotentials.
Pre-gauge freedom, parameters $B$, do not act on the analytic $h^{++}${--}prepotentials and can be used to gauge away all d.o.f. from the Mezincescu-type prepotential with the exception of the analytic prepotentials. All $B$-parameters are complex, except the parameters $B^{--}_{\alpha(s-1)\dot{\alpha}(s-1)}$ and $B^{--}_{\alpha(s-2)\dot{\alpha}(s-2)}$, which are real with respect to the tilde conjugation\footnote{Action of the tilde conjugation defined as $
		\widetilde{x^{\alpha\dot{\alpha}}} = x^{\alpha\dot{\alpha}},
		\widetilde{\theta^\pm_\alpha} = \bar{\theta}^\pm_{\dot{\alpha}},
		\widetilde{\bar{\theta}^\pm_{\dot{\alpha}}} = - \theta^\pm_\alpha,
		\widetilde{u^{\pm i}} = - u_i^\pm,
		\widetilde{u^\pm_i} = u^{\pm i}
		$; see~\cite{18}.}, namely,
\begin{equation}
	\widetilde{B^{--}_{\alpha(s-1)\dot{\alpha}(s-1)}}  = B^{--}_{\alpha(s-1)\dot{\alpha}(s-1)}, 
	\qquad
	\widetilde{B^{--}_{\alpha(s-2)\dot{\alpha}(s-2)}}  = B^{--}_{\alpha(s-2)\dot{\alpha}(s-2)}. 
\end{equation}	
Mezincescu-type prepotentials have rather complex non-geometric laws of the gauge transformation~\eqref{eq: Psi gauge and pre gauge}, which makes them less convenient for describing physical degrees of freedom in contrast to analytical prepotentials. Nevertheless, as we will see, they are more convenient for constructing the Abelian higher-spin cubic interactions and to determine equations on the corresponding $\mathcal{N}=2$ supercurrents. 

\section{Higher spin supercurrents: conservation equations}

Operating with the Mezincescu-type prepotentials $\Psi^{-\dots}$ and $\bar{\Psi}^{-\dots}$, one can construct interactions of the following form:
\begin{equation}\label{eq: Psi inter}
	S_{int} = \int d^4x d^8\theta du \; \left(\Psi^{-\alpha(s-1)\dot{\alpha}(s-2)} \mathcal{J}^+_{\alpha(s-1)\dot{\alpha}(s-2)}
	-
	\bar{\Psi}^{-\alpha(s-2)\dot{\alpha}(s-1)} \bar{\mathcal{J}}^+_{\alpha(s-2)\dot{\alpha}(s-1)}
	\right).
\end{equation}	
In order for the cubic vertices to be consistent, $\mathcal{N}=2$ supercurrents must satisfy the on-shell conservation laws. 
Gauge and pre-gauge transformations \eqref{eq: Psi gauge and pre gauge} imply the system of on-shell conservation equations:
		 \begin{equation}\label{eq: conservation supercurrents}
		 	\begin{cases}
			\mathcal{D}^{++} \mathcal{J}^+_{\alpha(s-1)\dot{\alpha}(s-2)} \approx 0,
			\\
			\mathcal{D}^{+\alpha} \mathcal{J}^+_{\alpha(s-1)\dot{\alpha}(s-2)} + \bar{\mathcal{D}}^{+\dot{\alpha}} \bar{\mathcal{J}}^+_{\alpha(s-2)\dot{\alpha}(s-1)} \approx 0,
			\\
			\mathcal{D}^+_{(\beta} \mathcal{J}^+_{\alpha(s-1))\dot{\alpha}(s-2)} \approx 0,
			\\
			\bar{\mathcal{D}}^+_{(\dot{\alpha}} \mathcal{J}^+_{\alpha(s-1)\dot{\alpha}(s-2))} - \mathcal{D}^+_{(\alpha} \bar{\mathcal{J}}^+_{\alpha(s-2))\dot{\alpha}(s-1)} \approx 0,
			\\
			\bar{\mathcal{D}}^{+\dot{\alpha}} \mathcal{J}^+_{\alpha(s-1)\dot{\alpha}(s-2)} \approx 0. 
			\end{cases}
	\end{equation}	
	Since we are constructing cubic vertices, these equalities must be fulfilled on the free equations of motion, presented in Section \ref{sec: 3}.


Before proceeding to the solution of the system of equations \eqref{eq: conservation supercurrents}, it is important to note that the structure of these equations implies that the solution has the special form
\begin{equation}\label{eq: J+ and J}
	\begin{cases}
	\mathcal{J}^+_{(\alpha_1\dots\alpha_{s-1})\dot{\alpha}(s-2)} = \mathcal{D}^+_{(\alpha_1} \mathcal{J}_{\alpha_2\dots\alpha_{s-1})\dot{\alpha}(s-2)},
	\\
	\bar{\mathcal{J}}^+_{\alpha(s-2)(\dot{\alpha}_1\dots\dot{\alpha}_{s-1})} = - \bar{\mathcal{D}}^+_{(\dot{\alpha}_1} \mathcal{J}_{\alpha(s-2)\dot{\alpha}_2\dots\dot{\alpha}_{s-1})};
	\end{cases}
\end{equation}	
where we introduced the real \textit{principle $\mathcal{N}=2$ supercurrent} $\mathcal{J}_{\alpha(s-2)\dot{\alpha}(s-2)}$, which satisfies the system of equations:
\begin{equation}\label{eq: primary supercurrent}
	\begin{cases}
		\widetilde{\mathcal{J}}_{\alpha(s-2)\dot{\alpha}(s-2)} = \mathcal{J}_{\alpha(s-2)\dot{\alpha}(s-2)},
		\\
		\mathcal{D}^{++} \mathcal{J}_{\alpha(s-2)\dot{\alpha}(s-2)} \approx 0,
		\\
		\mathcal{D}^{+\beta}\mathcal{J}_{(\beta\alpha_2\alpha_{s-2})\dot{\alpha}(s-2)} \approx 0,
		\\
		\bar{\mathcal{D}}^{+\dot{\beta}} \mathcal{J}_{\alpha(s-2)(\dot{\beta}\dot{\alpha}_2\dots\alpha_{s-2})} \approx 0.
	\end{cases}	
\end{equation}	
These conditions formally coincide with the definition of the $\mathcal{N}=2$ {\it primary supercurrent} \cite{Kuzenko:2021pqm}; see also~\cite{Howe:1981qj}, written in the harmonic superspace. However, since we are working with a non-conformal multiplets, the $\mathcal{N}=2$ superconformal properties that entail the laws of conservation in the article \cite{Kuzenko:2021pqm} are not fulfilled in our case. So these supercurrents are not consistent with conformal symmetry and we use alternative terminology.
In the case of $s=2$, one needs to replace the third and fourth equations by $(\mathcal{D}^+)^2 \mathcal{J} \approx 0$ and $(\bar{\mathcal{D}}^+)^2 \mathcal{J} \approx 0$, and it corresponds to the $\mathcal{N}=2$ supergravity supercurrent~\cite{Kuzenko:1999pi,Butter:2010sc}.

One can directly verify that supercurents \eqref{eq: J+ and J} satisfy the system of equations \eqref{eq: conservation supercurrents} due to equations \eqref{eq: primary supercurrent}. Hereby, equation \eqref{eq: J+ and J} implies that  $\mathcal{J}^+_{\alpha(s-1)\dot{\alpha}(s-2)}$, $\bar{\mathcal{J}}^+_{\alpha(s-2)\dot{\alpha}(s-1)}$ supercurrents can be considered as the descendants of the principle $\mathcal{N}=2$ supercurrent $\mathcal{J}_{\alpha(s-2)\dot{\alpha}(s-2)}$.
In terms of the principle $\mathcal{N}=2$ supercurrent, the cubic vertex \eqref{eq: Psi inter} has the form
\begin{equation}\label{eq: SF vertex}
	S_{int} = \int d^4xd^8\theta du \; \left( \Psi^{-\alpha(s-1)\dot{\alpha}(s-2)}\mathcal{D}^+_\alpha + \bar{\Psi}^{-\alpha(s-2)\dot{\alpha}(s-1)} \bar{\mathcal{D}}^+_{\dot{\alpha}} \right) \mathcal{J}_{\alpha(s-2)\dot{\alpha}(s-2)}.
\end{equation}	

Let us note that the principle $\mathcal{N}=2$ higher-spin supercurrents, as the consequence of \eqref{eq: primary supercurrent}, satisfy the conservation equation (for $s\geq 3$):
\begin{equation}
	\partial^{\beta\dot{\beta}} \mathcal{J}_{(\alpha(s-3)\beta)(\dot{\alpha}(s-3)\dot{\beta})} \approx 0.
\end{equation}	
Thus, all components of the principle $\mathcal{N}=2$ higher-spin supercurrent are conserved currents.

\section{Gauge-invariant higher spin supercurrents}\label{sec: 4}


\medskip

The general gauge-invariant anzatz for the principle spin $\mathbf{s}_1$ $\mathcal{N}=2$ higher-spin supercurrent, constructed from spin $\mathbf{s_2}$ superfields, has the form\footnote{We use only $\mathcal{N}=2$ Weyl-like higher-spin tensor, because other $\mathcal{N}=2$ higher-spin supercurvatures \cite{Ivanov:2024gjo} are zero on-shell and lead to "fake interactions".}: 
\begin{equation}
	\begin{split}
	\mathcal{J}_{\alpha(s_1-2)\dot{\alpha}(s_1-2)} =& \quad\sum_{p=0}^{s_1-2s_2} \alpha_p \; \partial^p \mathcal{W}_{\alpha(2s_2-2)} \partial^{s_1-2s_2-p} \bar{\mathcal{W}}_{\dot{\alpha}(2s_2-2)}
	\\&+ 
	\sum_{p=0}^{s_1-2s_2-1} \beta_p \; \partial^p \mathcal{D}^-_\alpha \mathcal{W}_{\alpha(2s_2-2)} \partial^{s_1-2s_2-p-1} \bar{\mathcal{D}}^+_{\dot{\alpha}}\bar{\mathcal{W}}_{\dot{\alpha}(2s_2-2)}
	\\&-
	\sum_{p=0}^{s_1-2s_2-1} \beta^*_{s_1-2s_2-p-1} \; \partial^p \mathcal{D}^+_\alpha \mathcal{W}_{\alpha(2s_2-2)} \partial^{s_1-2s_2-p-1} \bar{\mathcal{D}}^-_{\dot{\alpha}}\bar{\mathcal{W}}_{\dot{\alpha}(2s_2-2)}
	\\&+
	\sum_{p=0}^{s_1-2s_2-2} \gamma_p \;
	\partial^p \mathcal{D}^+_\alpha \mathcal{D}^-_\alpha \mathcal{W}_{\alpha(2s_2-2)} \partial^{s_1-2s_2-p-2} \bar{\mathcal{D}}^+_{\dot{\alpha}} \bar{\mathcal{D}}^-_{\dot{\alpha}} \bar{\mathcal{W}}_{\dot{\alpha}(2s_2-2)}.
	\end{split}
\end{equation}	
Here $\alpha_p$, $\beta_p$, and $\gamma_p$ are arbitrary complex constants, which can be determined from the system of equations \eqref{eq: primary supercurrent}. 
Bilinear supercurrents constructed from spin $\mathbf{s}_2$ superstrenghts minimally have $2s_2-2$ doted and $2s_2-2$ undoted indices and, therefore, can be coupled only to the spin $\mathbf{s_1}\geq 2\mathbf{s_2}$ supermultiplets.

Reality condition leads to the following constraints: 
\vspace{-0.2 cm}
\begin{equation}
		\tilde{\mathcal{J}}_{\alpha(s-2)\dot{\alpha}(s-2)}
		=
		\mathcal{J}_{\alpha(s-2)\dot{\alpha}(s-2)}
		\quad
		\Rightarrow
		\quad
		\begin{cases}
			\alpha_p = \alpha^*_{s_1-2s_2-p},
			\\
			\gamma_p = \gamma^*_{s_1-2s_2-p-2},
		\end{cases}		
\end{equation}	
whereas the harmonic independence condition gives the constraints to $\beta_p$ coefficients:
\begin{equation}
	\mathcal{D}^{++} 	\mathcal{J}_{\alpha(s-2)\dot{\alpha}(s-2)}
	=
	0
	\quad
	\Rightarrow
	\quad
	\beta_p = \beta^*_{s_1-2s_2-p-1},
\end{equation}	
while the conservation equation relates $\alpha_p$, $\beta_p$, and $\gamma_p$ coefficients:
\begin{equation}
	\mathcal{D}^{+\beta} \mathcal{J}_{\beta\alpha(s-3)\dot{\alpha}(s-2)} \approx 0
	\quad
	\Rightarrow
	\quad
	\begin{cases}
		\alpha_p (s_1-2s_2-p) + 4i \beta^*_{s_1-2s_2-p-1}(2s_2+p-1) = 0,
		\\
		\beta_p (s_1-2s_2-p-1) - 4i \gamma_p (2s_2+p) = 0.
	\end{cases}	
\end{equation}
All of them fix uniquely the coefficients $\alpha_p$, $\beta_p$, and $\gamma_p$
\begin{align}
\alpha_ p & = c \; (i)^{s_1-2s_2} (-1)^p \frac{\binom{s_1-2}{2s_2-2+p} \binom{s_1-2}{p}}{\binom{s_1-2}{2s_2-2}},
\qquad \qquad \,\, 
p = 0,1,2 \dots, s_1-2s_2, \\
\beta_p &= c\; (i)^{s_1-2s_2} (-1)^{p+1} \frac{1}{4i} \frac{\binom{s_1-2}{2s_2-1+p} \binom{s_1-2}{p}}{\binom{s_1-2}{2s_2-2}},
\qquad \,
p = 0,1,2 \dots, s_1-2s_2-1, \\
\gamma_p &= c\; (i)^{s_1-2s_2} (-1)^{p} \frac{1}{16} \frac{\binom{s_1-2}{2s_2+p} \binom{s_1-2}{p}}{\binom{s_1-2}{2s_2-2}},\qquad\qquad 
p = 0,1,2 \dots, s_1-2s_2-2
\end{align} 
up to the real constant $c$. 

	\enlargethispage{2\baselineskip} 
	
Let us comment on the obtained result. We know the structure of $\mathcal{N}=2$ superfield strengths only for the $\mathcal{N}=2$ higher-spin multiplets with integer highest spins \eqref{eq: SF strenghts}. However, the answer remains true for the half-integer supermultiplets if we assume that the corresponding $\mathcal{N}=2$ superfield strengths $\mathcal{W}_{\alpha(2s_2-1)}$ are known. In the limiting case $s_2=1$, the resulting $\mathcal{N}=2$ principle supercurrent coincides with the  $\mathcal{N}=2$ vector multiplet higher-spin primary supercurrent, previously obtained in \cite{Kuzenko:2023vgf}.
At the component level, the resulting $\mathcal{N}=2$ supercurrents are reduced to higher-spin currents of the form \eqref{eq: N=0 current}, and $\mathcal{N}=2$ higher-spin superfield vertex \eqref{eq: SF vertex} contains the cubic vertex \eqref{eq: type 2}.

\section{Discussion}\label{sec: dis}
\vspace{-0.2cm}
In this article we have constructed a new class of the consistent cubic interactions for $\mathcal{N}=2$ higher-spin gauge supermultiplets. It was shown, that $\mathcal{N}=2$ cubic vertex $(\mathbf{s_1}, \mathbf{s_2}, \mathbf{s_2})$ for $\mathbf{s_1}\geq 2\mathbf{s_2}$ has the universal structure:
\begin{equation}
	S_{int} = \int d^4xd^8\theta du \; \left( \Psi^{-\alpha(s_1-1)\dot{\alpha}(s_1-2)}\mathcal{D}^+_\alpha + \bar{\Psi}^{-\alpha(s_1-2)\dot{\alpha}(s_1-1)} \bar{\mathcal{D}}^+_{\dot{\alpha}} \right) \mathcal{J}_{\alpha(s_1-2)\dot{\alpha}(s_1-2)},
	\vspace{-0.2cm}
\end{equation}	
where $\Psi^-_{\alpha(s_1-1)\dot{\alpha}(s_1-2)}$, $\bar{\Psi}^-_{\alpha(s_1-2)\dot{\alpha}(s_1-1)}$ are the Mezincescu-type higher-spin prepotentials defined by \eqref{eq: Mez intr}, and $\mathcal{J}_{\alpha(s_1-2)\dot{\alpha}(s_1-2)}$ are  the principle gauge-invariant $\mathcal{N}=2$ supercurrents, which satisfy reality and harmonic independence conditions and the conservation equations $\mathcal{D}^{+\alpha} \mathcal{J}_{\alpha(s_1-2)\dot{\alpha}(s_1-2)} \approx 0$. We~also explicitly constructed spin $\mathbf{s_1}$ ($s_1\geq 2s_2$) $\mathcal{N} = 2$ higher-spin principle supercurrent for an arbitrary $\mathcal{N}=2$ off-shell gauge spin $\mathbf{s_2}$ supermultiplet 
\begin{equation}
	\begin{split}
		&\mathcal{J}_{\alpha(s_1-2)\dot{\alpha}(s_1-2)} = \;\; (i)^{s_1-2s_2} \sum_{p=0}^{s_1-2s_2}  (-1)^p \frac{\binom{s_1-2}{2s_2-2+p} \binom{s_1-2}{p}}{\binom{s_1-2}{2s_2-2}} \; \partial^p \mathcal{W}_{\alpha(2s_2-2)} \partial^{s_1-2s_2-p} \bar{\mathcal{W}}_{\dot{\alpha}(2s_2-2)}
		\\&
		- 
		(i)^{s_1-2s_2} \sum_{p=0}^{s_1-2s_2-1}   \frac{(-1)^p}{4i} \frac{\binom{s_1-2}{2s_2-1+p} \binom{s_1-2}{p}}{\binom{s_1-2}{2s_2-2}} \; \partial^p \mathcal{D}^-_\alpha \mathcal{W}_{\alpha(2s_2-2)} \partial^{s_1-2s_2-p-1} \bar{\mathcal{D}}^+_{\dot{\alpha}}\bar{\mathcal{W}}_{\dot{\alpha}(2s_2-2)}
		\\&
		+
			(i)^{s_1-2s_2} \sum_{p=0}^{s_1-2s_2-1} \frac{(-1)^p}{4i} \frac{\binom{s_1-2}{2s_2-1+p} \binom{s_1-2}{p}}{\binom{s_1-2}{2s_2-2}}  \; \partial^p \mathcal{D}^+_\alpha \mathcal{W}_{\alpha(2s_2-2)} \partial^{s_1-2s_2-p-1} \bar{\mathcal{D}}^-_{\dot{\alpha}}\bar{\mathcal{W}}_{\dot{\alpha}(2s_2-2)}
		\\&+
		(i)^{s_1-2s_2} \sum_{p=0}^{s_1-2s_2-2}   \frac{(-1)^{p}}{16} \frac{\binom{s_1-2}{2s_2+p} \binom{s_1-2}{p}}{\binom{s_1-2}{2s_2-2}}
		\partial^p \mathcal{D}^+_\alpha \mathcal{D}^-_\alpha \mathcal{W}_{\alpha(2s_2-2)} \partial^{s_1-2s_2-p-2} \bar{\mathcal{D}}^+_{\dot{\alpha}} \bar{\mathcal{D}}^-_{\dot{\alpha}} \bar{\mathcal{W}}_{\dot{\alpha}(2s_2-2)}
	\end{split}
\end{equation}	
in terms of the gauge-invariant $\mathcal{N}=2$ superfield strengths $\mathcal{W}_{\alpha(2s_2-2)}$ and $\bar{\mathcal{W}}_{\dot{\alpha}(2s_2-2)}$, which are expressed in terms of the analytic prepotentials, according to \eqref{eq: SF strenghts}.

\medskip

To construct vertices we used two types of $\mathcal{N} = 2$ higher-spin prepotentials: non-analytical Mezincescu-type and analytic prepotentials. Mezincescu-type prepotentials are connected to analytic prepotentials by the equations  \eqref{eq: G++ def} and \eqref{eq: G and Psi prepot}. The need to use non-analytical prepotentials is motivated by the non-analyticity of gauge-invariant $\mathcal{N}= 2$ superstrengths. In contrast to hypermultiplet higher-spin interactions in \cite{Buchbinder:2022kzl, Buchbinder:2022vra}, the straightforward construction of analytical gauge-invariant $\mathcal{N}=2$ supercurrents and corresponding couplings as integral over analytic superspace is not allowed. Perhaps analytical interactions can be constructed, e.g., by requiring the analyticity of supercurrents only on the equations of motion. We hope to explore this issue further.


\medskip

The obtained construction of the $\mathcal{N}=2$ higher-spin supercurrents opens up a number of interesting questions. The most natural of which are the construction of the higher-spin global and gauge transformations of the $\mathcal{N}=2$ actions \eqref{ActionsGen} and \eqref{eq: spin 1 action} corresponding to the constructed $\mathcal{N}=2$ supercurrents. These outcomes, the component analysis of the $\mathcal{N}=2$ supercurrents verticies, as well as the construction of the corresponding analytical $\mathcal{N}=2$ higher-spin supercurrents, will be presented in the separate paper \cite{IZ}.

\vspace{-0.4cm}

\acknowledgments

\vspace{-0.2 cm}

The author is indebted to Evgeny Ivanov for interest and valuable discussions.
The author  thanks Emmanouil  Raptakis and Konstantinos Koutrolikos for useful correspondence. 
The author is especially grateful to Polina Petriakova for her help in  the text preparation.
 Work was partially supported by the grant 22-1-1-42-2 from
the Foundation for the Advancement of Theoretical Physics and
Mathematics ``BASIS''.


\vspace{-0.4 cm}


\begin{thebibliography}{99}
	

	
	\bibitem{Ferrara:1974pz}
	S.~Ferrara and B.~Zumino,
	{\it Transformation Properties of the Supercurrent},
	\href{https://doi.org/10.1016/0550-3213(75)90063-2}{Nucl. Phys. B \textbf{87} (1975), 207}.
	

	
	\bibitem{Ogievetsky:1976qc}
	V.~Ogievetsky and E.~Sokatchev,
	{\it On Vector Superfield Generated by Supercurrent},
	\href{https://doi.org/10.1016/0550-3213(77)90318-2}{Nucl. Phys. B \textbf{124} (1977), 309-316}.
	

	\bibitem{Howe:1981qj}
	P.~S.~Howe, K.~S.~Stelle and P.~K.~Townsend,
	{\it Supercurrents},
	\href{https://doi.org/10.1016/0550-3213(81)90429-6}{Nucl. Phys. B \textbf{192} (1981), 332-352}.
	

	
	\bibitem{Kuzenko:1999pi}
	S.~M.~Kuzenko and S.~Theisen,
	\textit{Correlation functions of conserved currents in $\mathcal{N}=2$ superconformal theory},
	\href{https://doi.org/10.1088/0264-9381/17/3/307}{Class. Quant. Grav. \textbf{17} (2000), 665-696}
	[arXiv:hep-th/9907107 [hep-th]].
	
	\enlargethispage{2\baselineskip}
	
	\bibitem{Butter:2010sc}
	D.~Butter and S.~M.~Kuzenko,
	{\it $\mathcal{N}=2$ supergravity and supercurrents},
	\href{https://doi.org/10.1007/JHEP12(2010)080}{JHEP \textbf{12} (2010), 080}
	[arXiv:1011.0339 [hep-th]].
	

	
		\bibitem{18} A.~S.~Galperin, E.~A.~Ivanov, V.~I.~Ogievetsky, E.~S.~Sokatchev,
	{\it Harmonic superspace}, Cambridge Monographs on Mathematical
	Physics, Cambridge University Press, 2001, 306 p.
	
	\bibitem{Smirnov:2013kba}
	P.~A.~Smirnov and M.~A.~Vasiliev,
	{\it Gauge-noninvariant higher-spin currents in four-dimensional Minkowski space},
	\href{https://doi.org/10.1007/s11232-014-0231-5 }{Theor. Math. Phys. \textbf{181} (2014) no.3, 1509-1521}
	[arXiv:1312.6638 [hep-th]].
	
	\bibitem{Berends:1985xx}
	F.~A.~Berends, G.~J.~H.~Burgers and H.~van Dam,
	{\it Explicit Construction of Conserved Currents for Massless Fields of Arbitrary Spin},
	\href{https://doi.org/10.1016/S0550-3213(86)80019-0}{Nucl. Phys. B \textbf{271} (1986), 429-441}.

	\bibitem{Bel}
	L. Bel,
	{\it Introduction d’un tenseur du quatri`eme ordre},
	Acad. Sci. Paris, Comptes Rend. 248,
	1297 (1959).
	
	\bibitem{Deser:1990bk}
	S.~Deser and Z.~Yang,
	{\it Inconsistency of Spin 4 - Spin-2 Gauge Field Couplings},
	\href{https://doi.org/10.1088/0264-9381/7/8/024}{Class. Quant. Grav. \textbf{7} (1990), 1491-1498}.


	\bibitem{Manvelyan:2009vy}
	R.~Manvelyan, K.~Mkrtchyan and W.~Ruhl,
	{\it Off-shell construction of some trilinear higher spin gauge field interactions},
	\href{https://doi.org/10.1016/j.nuclphysb.2009.07.007}{Nucl. Phys. B \textbf{826} (2010), 1-17}
	[arXiv:0903.0243 [hep-th]].
	



	
	

	
	

		
	
	
	
	
	\bibitem{Buchbinder:2018wzq}
	I.~L.~Buchbinder, S.~J.~Gates and K.~Koutrolikos,
	{\it Conserved higher spin supercurrents for arbitrary spin massless supermultiplets and higher spin superfield cubic interactions},
	\href{https://doi.org/10.1007/JHEP08(2018)055}{JHEP \textbf{08} (2018), 055}
	[arXiv:1805.04413 [hep-th]].
	
	
	\bibitem{Gates:2019cnl}
	S.~J.~Gates and K.~Koutrolikos,
	{\it Progress on cubic interactions of arbitrary superspin supermultiplets via gauge invariant supercurrents},
	\href{https://doi.org/10.1016/j.physletb.2019.134868}{Phys. Lett. B \textbf{797} (2019), 134868}
	[arXiv:1904.13336 [hep-th]].
	

	
	\bibitem{Galperin:1984av}
	A.~Galperin, E.~Ivanov, S.~Kalitzin, V.~Ogievetsky and E.~Sokatchev,
	{\it Unconstrained $\mathcal{N}=2$ Matter, Yang-Mills and Supergravity Theories in Harmonic Superspace},
	\href{https://doi.org/10.1088/0264-9381/1/5/004}{Class. Quant. Grav. \textbf{1} (1984), 469-498}
	[erratum: Class. Quant. Grav. \textbf{2} (1985), 127].
	

			
		\bibitem{Buchbinder:2021ite}
		I.~Buchbinder, E.~Ivanov and N.~Zaigraev,
		\textit{Unconstrained off-shell superfield formulation of 4D, $ \mathcal{N} $ = 2 supersymmetric higher spins},
		\href{https://doi.org/10.1007/JHEP12(2021)016}{JHEP \textbf{12} (2021), 016}
		[arXiv:2109.07639 [hep-th]].
		
	
		
		
		\bibitem{Buchbinder:2022kzl}
		I.~Buchbinder, E.~Ivanov and N.~Zaigraev,
		{\it Off-shell cubic hypermultiplet couplings to $ \mathcal{N} $ = 2 higher spin gauge superfields},
		\href{https://doi.org/10.1007/JHEP05(2022)104}{JHEP \textbf{05} (2022), 104}
		[arXiv:2202.08196 [hep-th]].
		
		
	
		
		
		\bibitem{Buchbinder:2022vra}
		I.~Buchbinder, E.~Ivanov and N.~Zaigraev,
		{\it $ \mathcal{N} $ = 2 higher spins: superfield equations of motion, the hypermultiplet supercurrents, and the component structure},
		\href{https://doi.org/10.1007/JHEP03(2023)036}{JHEP \textbf{03} (2023), 036}
		[arXiv:2212.14114 [hep-th]].
		
		
		\bibitem{Buchbinder:2024pjm}
		I.~Buchbinder, E.~Ivanov and N.~Zaigraev,
		{\it $\mathcal{N} = 2$ superconformal higher-spin multiplets and their hypermultiplet couplings},
		\href{https://doi.org/10.1007/JHEP08(2024)120}{JHEP \textbf{08} (2024), 120}
		[arXiv:2404.19016 [hep-th]].
		
		\bibitem{Zaigraev:2023ogo}
		N.~Zaigraev,
		{\it $\mathcal{N} = 2$ Higher Spin Theory in Harmonic Superspace},
		\href{https://doi.org/10.1134/S1063779623060254}{Phys. Part. Nucl. \textbf{54} (2023) no.6, 1084-1088}.
		
		\bibitem{Zaigraev:2024xve}
		N.~Zaigraev, I.~Buchbinder and E.~Ivanov,
		{\it $\mathcal{N} = 2$ higher spin theories and harmonic superspace},
		\href{https://doi.org/10.22323/1.455.0048}{PoS \textbf{ICPPCRubakov2023} (2024), 048}
		[arXiv:2402.05704 [hep-th]].
		
		
	
		
		\bibitem{Kuzenko:2023vgf}
		S.~M.~Kuzenko and E.~S.~N.~Raptakis,
		{\it On higher-spin $ \mathcal{N} $ = 2 supercurrent multiplets},
		\href{https://doi.org/10.1007/JHEP05(2023)056}{JHEP \textbf{05} (2023), 056}
		[arXiv:2301.09386 [hep-th]].
		
		\bibitem{Kuzenko:2021pqm}
		S.~M.~Kuzenko and E.~S.~N.~Raptakis,
		{\it Extended superconformal higher-spin gauge theories in four dimensions},
		\href{https://doi.org/10.1007/JHEP12(2021)210}{JHEP \textbf{12} (2021), 210}
		[arXiv:2104.10416 [hep-th]].
		
	\bibitem{Ivanov:2024gjo}
	E.~Ivanov and N.~Zaigraev,
	{\it Off-shell invariants of linearized $4D$,$\mathcal{N}=2$ supergravity in the harmonic approach},
	\href{https://doi.org/10.1103/PhysRevD.110.066020}{Phys. Rev. D \textbf{110} (2024) no.6, 066020}
	[arXiv:2407.08524 [hep-th]].
		
		\bibitem{Mezincescu}
		L. Mezincescu, 
		{\it On the superfield formulation of O(2) supersymmetry}, Dubna preprint JINR-P2-12572 (June, 1979).
		
		\bibitem{Gates:1981qq}
		S.~J.~Gates, Jr. and W.~Siegel,
		{\it Linearized  $\mathcal{N}=2$ superfield supergravity},
		\href{https://doi.org/10.1016/0550-3213(82)90047-5}{Nucl. Phys. B \textbf{195} (1982), 39-60}.
		\enlargethispage{2\baselineskip} 
		\bibitem{Zupnik:1998td}
		B.~M.~Zupnik,
		{\it Background harmonic superfields in $\mathcal{N}=2$ supergravity},
		\href{https://doi.org/10.1007/BF02557138}{Theor. Math. Phys. \textbf{116} (1998), 964-977}
		[arXiv:hep-th/9803202 [hep-th]].

		\bibitem{IZ}
		E. Ivanov and N. Zaigraev,
		\textit{work in progress}.


		\bibitem{Kuzenko:2024vms}
		S.~M.~Kuzenko and E.~S.~N.~Raptakis,
		{\it Towards $\mathcal{N}=2$ superconformal higher-spin theory},
		[arXiv:2407.21573 [hep-th]].
	


\end{thebibliography}
\end{document}